\documentclass[conference]{IEEEtran}
\IEEEoverridecommandlockouts
\usepackage{cite}
\usepackage{amsmath,amssymb,amsfonts}
\usepackage{algorithmic}
\usepackage{graphicx}
\usepackage{textcomp}
\usepackage{xcolor}
\usepackage{booktabs} 
\def\BibTeX{{\rm B\kern-.05em{\sc i\kern-.025em b}\kern-.08em
    T\kern-.1667em\lower.7ex\hbox{E}\kern-.125emX}}
\begin{document}

\title{Advanced Anomaly Detection and Threat Intelligence in Zero Trust IoT Environments Using Machine Learning}

\author{\IEEEauthorblockN{Muhammad Umair Basharat\IEEEauthorrefmark{1}, 
Jawad Hussain\IEEEauthorrefmark{1}, 
Waqas Khalid\IEEEauthorrefmark{2}, 
Chiew Foong Kwong\IEEEauthorrefmark{2}}
\IEEEauthorblockA{\IEEEauthorrefmark{1}School of Information Technology, Whitecliffe, Auckland, New Zealand}
\IEEEauthorblockA{\IEEEauthorrefmark{2}Department of Electrical \& Electronic Engineering, University of Nottingham Ningbo China (UNNC), Ningbo, China}
}

\maketitle

\begin{abstract}
The growing adoption of IoT and cloud computing, combined with rapid advancements in digital technologies, has considerably increased the cyber-attack surface, resulting in increasingly complex and persistent attacks. Traditional security methods, primarily based on perimeter defenses, are insufficient to meet these developing threats, especially within the context of a Zero Trust Security (ZTS) architecture. This study investigates the application of sophisticated artificial intelligence (AI) and machine learning (ML) techniques, including the use of the Synthetic Minority Oversampling Technique (SMOTE), to improve anomaly detection and threat intelligence systems. This study focuses on how Support Vector Machine (SVM), Random Forest (RF), and Decision Tree (DT) classifiers might increase threat detection accuracy in IoT environments. The research endeavors to improve cybersecurity resilience by mitigating false positives and providing actionable intelligence through supervised learning algorithms. The KDD Cup 1999 dataset is used in the study to assess how well these models perform in simulating various network intrusions and regular traffic. The application of SMOTE significantly enhanced the performance of these models by addressing class imbalance, leading to improved detection accuracy. Furthermore, as supplementary methods for detecting malicious URLs and advanced persistent threats (APTs), edge-based machine learning and blockchain technology are investigated. This study addresses the shortcomings of conventional security systems and supports the growing demand for reliable threat detection in a world that is becoming more interconnected. It also advances the creation of more proactive and adaptable cybersecurity solutions.
\end{abstract}


\section{Introduction}
\subsection{Problem Statement}
Advanced Persistent Threats (APTs) highlight flaws in current detection methods because these lengthy, complex, and covert assaults are often carried out by skilled adversaries and can remain undetected while causing strategic harm [1, 2], and their ability to blend with regular network traffic makes it hard to detect using traditional systems, which struggle to identify subtle patterns in user or system activity [3], while these methods also fail to detect advanced anomalies, such as multistage attacks or lateral network movement, leaving organizations vulnerable to significant infrastructure breaches and data loss [4], therefore incorporating artificial intelligence (AI) and machine learning (ML)-based technologies into the Zero Trust Security architecture is essential to overcome these constraints [5, 6], as adaptive and intelligent anomaly detection systems powered by AI and ML can learn from dynamic threats and network activity in real-time and quickly analyze large datasets to identify trends, patterns, and outliers that might signal vulnerabilities, thereby reducing false positives and enhancing precision in identifying known and unknown threats [7], while AI-powered threat intelligence systems further bolster security by offering actionable insights beyond traditional log analysis through correlation and contextualization [8, 9], enabling organizations to foresee vulnerabilities and take preventative measures through real-time threat intelligence [9, 10], and this study aims to explore the creation and application of AI and ML technologies to enhance anomaly detection and threat intelligence within Zero Trust frameworks, particularly in IoT networks, by addressing challenges such as identifying complex attacks like APTs, reducing false positives, and ensuring scalability in distributed systems, so that integrating AI-driven anomaly detection with Zero Trust principles may advance cybersecurity and influence future infrastructure designs to be more robust and proactive [11, 12].
\subsection{Background}
The prevalence of Internet of Things (IoT) devices has transformed numerous sectors over the past years, including automating processes, making decisions based on the data, and facilitating the communication process [13]. The IoT creation provides unprecedented comfort and effectiveness in the domains of smart homes, healthcare, transportation, and industrial automation [10]. Nevertheless, organizations are finding it difficult to secure their networks as IoT networks grow at a high rate because of the numerous connected and geographically distributed devices that attract cyberattacks [2, 5]. One of the most significant aspects of the security of the IoT is the limited power and processing resources of most devices that restrict the ability to implement effective protection systems and lead to the lack of overall security levels in heterogeneous environments [14]. Conventional security tools, including firewalls and rule-based intrusion detection systems, do not always succeed in such dynamic environments and may fail to identify device-level threats, network-level threats, and data breaches, including firmware manipulation, DDoS attacks, and data breaches [3, 4].

Threat intelligence and anomaly detection are thus of high importance in enhancing IoT cybersecurity through detection of unusual behavior and proactive protection [4, 9]. Complex machine learning approaches, like Random Forest (RF), Support Vector Machine (SVM), and Decision Tree (DT), enhance the level of detection by identifying complex patterns based on massive datasets and addressing changes in threats [15]. At the intersection of AI- and ML-based detection systems, Zero Trust Security (ZTS) schemes, which involve the constant verification of trust, further increase the security of the distributed IoT setting [5, 6]. This study aims to address the weaknesses of the existing models and help create more robust, smart, and active IoT security levels by integrating AI-based anomaly detection with Zero Trust principles [10, 11].
\subsection{Objectives}
\subsubsection{Research Objective}
The primary objective of this research is to enhance detection accuracy in Internet of Things (IoT) environments using advanced machine learning models [5]. The study focuses on three key models: Random Forest (RF), Support Vector Machine (SVM), and Decision Tree (DT) [15]. These models will be developed and assessed using the KDD Cup 1999 dataset, which provides a comprehensive range of network intrusion data [15]. The goals include improving the detection of anomalies and cyber threats, reducing false positives, and strengthening online security [3, 4]. The project aims to deliver more proactive and efficient security solutions for IoT networks by leveraging the interpretability of DT, the resilience of SVM in high-dimensional spaces, and the feature significance of RF [15].
\subsubsection{Research Questions}
\begin{itemize}
\item How can AI and ML be used to improve anomaly detection and threat intelligence in Zero Trust security frameworks?
\item What are the most effective AI and ML techniques for detecting advanced cyber threats, and how can they be integrated into Zero Trust Security?
\item How can combining SVM and RF improve the accuracy of intrusion detection systems in Zero Trust Security?
\item In what ways can ensemble learning models improve the detection of malicious URLs and strengthen web security in IoT networks within a Zero Trust framework?
\end{itemize}
\subsubsection{Significance}
This study contributes by positioning AI- and ML-driven anomaly detection and threat intelligence as practical enablers of Zero Trust Security in IoT environments, where traditional perimeter and signature-based defenses struggle to scale and adapt [4, 6, 7]. First, it supports improved detection reliability by evaluating supervised learning models (RF, SVM, and DT) for intrusion/anomaly detection and emphasizing approaches that reduce false positives and missed detections, key operational barriers in real deployments [15]. Second, it advances the integration of threat intelligence into Zero Trust decision-making by emphasizing correlation and context-aware analysis, which strengthens proactive defense against sophisticated and persistent attacks [8, 9]. Third, by considering IoT constraints (heterogeneity, limited resources, and distributed operation), the study informs scalable and resource-aware security design choices aligned with modern Zero Trust architecture evolution [5].

\section{Literature Review}

\subsection{Traditional Security Models vs. Zero Trust Security}
Conventional security approaches rely heavily on firewalls, intrusion detection systems, and rule-based filtering mechanisms to protect network perimeters [16]. These models assume that entities inside the network are trustworthy, an assumption that is increasingly invalid in modern distributed systems where access is dynamic and credentials can be compromised [3, 11]. As networks incorporate IoT devices, cloud services, and remote access infrastructures, attack surfaces expand and perimeter boundaries dissolve, weakening the effectiveness of perimeter-centric enforcement points [2, 12]. Zero Trust Security frameworks address these weaknesses by enforcing continuous trust verification, micro-segmentation, and least-privilege access controls, thereby limiting implicit trust and restricting lateral movement [6]. Research also shows Zero Trust is being adapted across industrial and sector-specific contexts, including industrial IoT and smart manufacturing, which further demonstrates its relevance to distributed cyber-physical infrastructures [12]. However, while architectural principles are well established, effective real-time detection mechanisms embedded within ZTS environments remain an open challenge, particularly when trust signals must be generated from diverse telemetry sources at scale [6, 8].

\subsection{AI and ML in Anomaly Detection}
Artificial intelligence (AI) and machine learning (ML) have become the new revolution in detecting anomalies because of the restriction of the traditional rule-based system that cannot cope with the zero-day behaviour and the change of attack patterns [4, 7]. Models of supervised learning trained on labelled data have also exhibited superior detection accuracy and fewer false positives in dynamic settings and where features are able to detect behavioural deviation, but not signatures in isolation [3, 15]. RF models are resistant to overfitting and give insight into the importance of features, which are useful in determining important indicators of an anomaly and is valuable to operational security tuning and interpretability of the model [15, 17]. The Support Vector Machines (SVM) can be beneficial in the case of binary classification, optimization of the separation between the normal and non-normal patterns, but the performance is determined by the choice of the kernel and parameter optimization.

Other algorithms such as K-Nearest Neighbors (KNN) offer simplicity but face scalability and latency constraints in high-dimensional or large-volume traffic data [3, 18], while Gradient Boosting improves pattern recognition in complex datasets through iterative error minimization and can capture subtle attack behaviors. The use of ensemble methods also increases the reliability of detection through the combination of different classifiers, which increases predictive performance and robustness to mixed traffic patterns [19]. Simultaneously, explainability has also become a design concern since models that perform well but cannot explain the alerts can decrease the confidence of the analysts and can obstruct its implementation in the security operations procedures [4, 20]. Although these advances have been made, the issues of computational overhead, concept drift, and real-time adaptation are still imperative in deployed IoT security systems [5, 7].

\subsection{Cyber Threat Intelligence and Detection Accuracy}
Cyber threat intelligence (CTI) coupled with AI-based models improves contextual awareness and predictive protection qualities through the correlation of indicators, behavioral cues, and multi-source telemetry to aid in the proactive detection of risk [8, 9]. According to CTI research, actionable intelligence is also based on how well the signals have been shared, interpreted, and operationalized into defensive controls, as well as detection [8]. Clustering with optimization rate classifiers are among the methods that have been used in detecting malicious URLs and network anomalies, which enhance early detection of web-based threats that have the potential to impact the IoT ecosystems and appearance services [19, 21]. However, elevated levels of false positive are still overwhelming the security operations centers, lowering the efficiency of the analysts, and augmenting the risk of alert fatigue in which important threats could be missed [3, 4]. Ensemble learning enhances the reliability of detection at a high cost in terms of the complexity of computations, which requires compromise between efficiency and precision, particularly when deployed in edge or hybrid IoT designs [19].

\section{Methodology and Ethics}

\subsection{Research Design and Approach}
This study adopts a quantitative research design focused on numerical evaluation of machine learning models for anomaly and intrusion detection in IoT-like network environments [3], [4]. The study is both experimental and comparative in nature because several supervised classifiers would be trained and evaluated in the same conditions to compare the results in terms of performance [15]. This method is appropriate since the research questions demand quantifiable measures of model efficiency by standardized measures of such quantities as accuracy, precision, recall, and F1-score [4].

\subsection{Participants, Subjects, or Data Sources}
The study does not involve human participants; instead, it uses a benchmark intrusion detection dataset as the primary data source. The KDD Cup 1999 data is chosen due to the purposive sampling strategy, as it consists of labeled normal and attack traffic and is generally used to benchmark intrusion detection [15, 17]. Sample size is the records of the dataset after preprocessing, and only valid instances of network connection that are fully labelled are included and corrupted, duplicate, and unusable records are removed in the cleaning process [4].

\subsection{Data Collection Methods and Tools}
Data is collected from the KDD Cup 1999 repository and imported into Python for analysis [17]. The tools used include Python with pandas and NumPy for data handling, and scikit-learn for preprocessing, model training, and evaluation [3]. The procedure includes (1) dataset loading, (2) categorical encoding using Label Encoder, (3) feature scaling using StandardScaler where required, (4) train–test splitting, and (5) class balancing using SMOTE to address minority-class underrepresentation [4]. The study is conducted in a controlled offline computing environment, and the dataset-based approach avoids external environmental variation that could influence results.

\subsection{Data Analysis Methods}
Machine learning classification modelling and metric-based evaluation Python is utilized to analyze data [3]. The considered models include Decision Tree (DT), Random Forest (RF), and Support Vector Machine (SVM), as they have been used in the context of intrusion detection previously and feature strengths and weaknesses that complement each other [15, 22]. Data preparation involves the cleaning of duplicates, the encoding of the categorical variables, scaling the numeric features, and the imbalance with SMOTE [4]. The metrics of model performance are accuracy, precision, recall, and F1-score, where F1-score is more important due to the imbalance in intrusion detection datasets and the misleading effect of accuracy [3, 4].

\subsection{Ethical Considerations}
The IRB approval is not needed as the use of the non-human, public dataset does not involve the gathering of personal and sensitive data of the participants. The ethical practice is upheld through the transparent reporting of limitations of the dataset and the absence of misleading findings on the basis of benchmark results [4]. The risks of privacy are low since KDD Cup 1999 is anonymized, publicly available to research, and the findings are presented in a responsible manner since false positives and false negatives may impact the actual security activities in the field once rolled out without validation [3].

\subsection{Limitations}
A key limitation is reliance on the KDD Cup 1999 dataset, which may not represent modern IoT traffic patterns or emerging threats, potentially affecting real-world generalization [3]. Another limitation is that the study evaluates models in a controlled dataset environment rather than live IoT systems, so latency, resource constraints, and operational factors may differ in practice [5]. Additionally, SMOTE may improve minority-class learning but can introduce synthetic patterns that may not perfectly reflect real attack distributions [4].

\subsection{Reliability and Validity}
Reliability is ensured by using a consistent pipeline of experimentation, standardized preprocessing, and applying common metrics of benchmark evaluation to all the models [15]. Generalization is also evaluated through the train-test splitting, which validates the intrusion dataset and models of intrusion testing on unknown data [3, 17]. Construct validity is analyzed by measuring the evaluation metrics (precision, recall, F1) with intrusion detection purposes, particularly when the attack is unbalanced [4].

\section{Results and Analysis}

\subsection{Experimental Evaluation Setup}
The proposed anomaly detection framework was evaluated using the KDD Cup 1999 benchmark dataset. Three supervised machine learning classifiers Decision Tree (DT), Random Forest (RF), and Support Vector Machine (SVM) were implemented using Scikit-learn. Deep Learning (DL) and Recurrent Neural Network (RNN) models were also evaluated for comparative analysis. To address class imbalance in the dataset, the Synthetic Minority Oversampling Technique (SMOTE) was applied prior to training. Performance was assessed using Accuracy and weighted F1-score to ensure balanced evaluation across attack categories.

\subsection{Baseline Model Performance}
Table \ref{tab:performance} summarizes Train Score, Test Score, and F1-score across models. The strongest overall results were achieved by Random Forest with SMOTE, attaining near-perfect predictive capability with a minimal generalization gap. In comparison, Random Forest (baseline) achieved strong results, demonstrating that SMOTE provided a measurable improvement, particularly in minority-class representation reflected in F1.

The Decision Tree recorded strong overall performance but a larger train–test separation, which indicates higher variance and potential sensitivity to training patterns. The SVM produced lower overall accuracy but a comparatively high F1-score, suggesting that it detected minority events more consistently than accuracy alone would imply. The deep learning model yielded stable results, reflecting good generalization, but did not surpass ensemble methods. The RNN achieved moderate accuracy but an extremely low F1 score, meaning the model largely failed to identify the minority class correctly, consistent with weak sequence structure in the underlying tabular features.
\begin{table}[htbp]
\caption{Model Performance Comparison}
\centering
\begin{tabular}{lccc}
\toprule
\textbf{Model} & \textbf{Train Score} & \textbf{Test Score} & \textbf{F1 Score} \\
\midrule
Decision Tree & 0.996314 & 0.991135 & 0.996035 \\
Random Forest & 0.998247 & 0.995663 & 0.996159 \\
SVM & 0.942724 & 0.943636 & 0.965913 \\
Deep Learning & 0.968527 & 0.971818 & 0.968515 \\
RNN & 0.957015 & 0.954882 & 0.206943 \\
SMOTE Random Forest & 0.999734 & 0.996902 & 0.996902 \\
\bottomrule
\end{tabular}
\label{tab:performance}
\end{table}

\begin{figure}[htbp]
\centerline{\includegraphics[width=0.7\columnwidth]{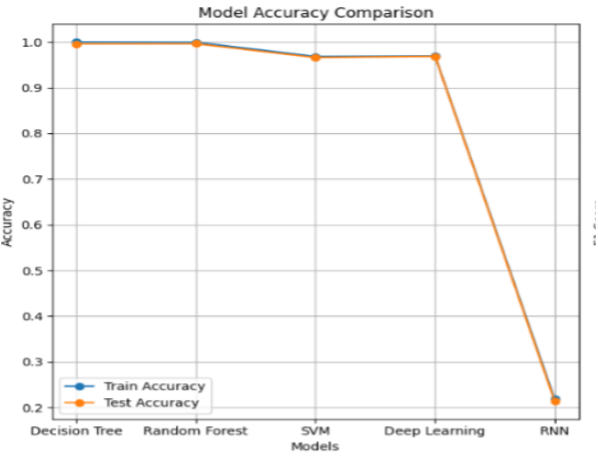}} 
\caption{Train vs Test Accuracy Comparison}
\label{fig:accuracy}
\end{figure}
The Random Forest classifier achieved the highest test accuracy (99.56\%) among all baseline models as shown in Fig.~\ref{fig:accuracy}. Decision Tree demonstrated near-perfect training accuracy, indicating slight overfitting. SVM exhibited stable but lower performance compared to ensemble-based methods.

Although deep learning models achieved competitive accuracy, they did not surpass Random Forest. The RNN model showed a significantly low F1-score, indicating poor minority-class detection in tabular intrusion data.

\subsection{Impact of SMOTE on Detection Quality}
The presence of class imbalance may generate false high accuracy in intrusion detection, since it is possible to predict the majority class many times, inflating the accuracy and missing attacks. In this experiment, SMOTE boosted reliability in identifying minority-classes and the greatest improvement was witnessed in Random Forest. Although baseline Random Forest had already been quite a strong performer, SMOTE increased the test score by 0.995663 to 0.996902, and it also matched the F1-score with accuracy, which showed less misses of the minority-class.

In the case of Decision Trees, SMOTE was better at achieving better detection balance, but not completely eliminate overfitting behaviours, which is probably because individual tree learners may draw very narrow boundaries that do not always form when applied to unseen samples. In the case of SVM, SMOTE enhanced the sensitivity of the minority classes (high F1) although the overall accuracy was less implying that the margin-based decision boundary was not able to learn the multi-class structure as well as it did in ensemble trees.

\subsection{Comparative Efficiency and Practicality for IoT Deployment}
Among all evaluated models, Random Forest consistently outperformed DT, SVM, DL, and RNN in both accuracy and F1-score. The superior performance of RF can be attributed to:
\begin{enumerate}
\item Ensemble-based variance reduction through bagging
\item Robust handling of high-dimensional features
\item Feature importance-based decision aggregation
\end{enumerate}
Decision Trees provided high interpretability but exhibited slight overfitting. SVM showed stable generalization, but lower classification accuracy compared to RF.

Deep learning models demonstrated scalability but did not outperform tree-based ensemble methods for structured intrusion datasets. RNNs, designed primarily for sequential data, were less effective for static network traffic features.

Regarding the deployment, the best trade-off among accuracy, robustness and feasibility is the case of Random Forest with SMOTE. Its ensemble nature minimizes variance, enhances stability in noisy settings, and enables to operate with strong performance without the need to have deep architectures.

\begin{itemize}
\item Random Forest (SMOTE): Most suitable for gateway/edge deployment and centralized SOC pipelines, due to high accuracy, strong generalization, and robustness.
\item Decision Tree: Highly interpretable and computationally light, suitable for simple edge implementations, but less stable across sampling variations; pruning/depth control is necessary.
\item SVM: Generally stable and effective in high-dimensional contexts, but in this dataset, it underperformed relative to tree ensembles, suggesting limited suitability for heterogeneous intrusion patterns.
\item Deep Learning: Good generalization but higher compute cost; more practical at gateways or cloud layers than endpoints.
\item RNN: Not recommended for this representation; sequence models require meaningful temporal ordering and sequential telemetry.
\end{itemize}

\section{Discussion}

\subsection{Interpretation of Findings}
These findings suggest that tree-based ensemble learning is the most effective in intrusion/anomaly detection when the imbalance in the classes is explicitly addressed. The better results of the SMOTE Random Forest indicate that an equal exposure to the minority and the majority class enhances the voting process of the ensemble and its predictive capability in realistic variations [3, 4]. This coincides with the literature of IoT anomaly detection indicating that ML techniques are superior to traditional techniques in noisy, heterogeneous IoT traffic where attack patterns do not always occur at the same devices and places [5].

The Decision Tree had a high performance but was more sensitive to training patterns, as has always been the case with single trees which tend to overfit when the decision boundaries become over-specific [3]. The SVM scores indicate more stable learning and less overall class separation, and the margin-based optimization can be less expressive than the ensemble trees in heterogeneous intrusion feature spaces [22]. Deep learning performed well and gave consistent results, though it was not better than ensemble methods, which is also in line with the literature that deep models do not necessarily perform out of strong ensemble baselines on structured intrusion data [23]. The fact that the F1 of the RNN was low supports the fact that deep sequence models should be combined with meaningful temporal data representation to detect patterns of minority-classes [23].

\subsection{Comparison with Traditional and Existing Techniques}
ML-based detection has better aptitude to identify novel or developing attacks as compared to traditional signature-based IDS approaches as the model learns discriminative relationships as opposed to using rigid patterns [5]. Ensemble learning is widely recognized as a robust approach in anomaly detection because it reduces variance and improves stability under noisy, multi-feature conditions [3].

The findings of the Random Forest can be compared with the known studies on intrusion detection results, which show that ensemble classifier effectively uses a network security setting [3, 15]. In addition, the post-SMOTE improvement is a further argument in favor of an important idea in anomaly detection study, namely, intrusion datasets are frequently asymmetric, and accuracy alone may mask poor minority-class performance [4]. Minority-class detection is frequently more important than the overall accuracy in the context of operational security as false negatives may result in breaches and lateral movement in the IoT environment [5].

\subsection{Novelty, Strengths, and Limitations}
The paper unites the ML-based intrusion detection with explicit imbalance correction and frames the results in the context of the Zero Trust principles, where constant verification is based on behavioral risk indicators, and no trust is presumed [6, 11].

The SMOTE Random Forest showed the best overall performance, which supports the benefits of the ensemble learning in intrusion detection [15]. F1-score is utilized as one of the fundamental assessment measures, which enhances validity in case of class imbalance [4]. The framework can also be directly applied to IoT and Zero Trust environments where adaptive access control decisions can use detection outputs to make decisions [6].

The KDD-based dataset lacks full representation of modern encrypted IoT telemetry or current attack pattern, which is very common in the literature related to IoT anomaly detection [5]. The fact that the RNN is not performing well demonstrates the significance of well-structured temporal data to sequence-based models [23]. Moreover, in comparison to the Decision Trees, the Random Forest models are not as interpretable as Decision Trees, and explainable AI methods might be needed to enhance the trust of the analyst and compliance with governance principles [20]. Direct deployment can also be constrained by resource limitations on the use of the limited endpoints of IoT [7].

\subsection{Implications for Zero Trust Security in IoT}
Zero Trust focuses on constant checks and access decisions based on risk as opposed to trust on the perimeter [11], [6]. The high accuracy of the SMOTE Random Forest model indicates that anomaly detection methods based on the use of ML can be an effective risk-scoring element in the Zero Trust pipelines. The presence of precise anomaly detection allows the detection of suspicious sessions early, as well as adaptive controls, e.g., segmentation or step-up authentication [6].

Since the effectiveness of Zero Trust enforcement relies on the quality of telemetry and behavioral monitoring, a more efficient process of anomaly detection directly increases the effectiveness of Zero Trust in IoT environment, in which the identity and behaviors of devices cannot be constant [6].

\section{Conclusion and Recommendations}

\subsection{Conclusion}
This study evaluated three supervised machine learning models Random Forest enhanced with SMOTE, Support Vector Machine (SVM), and Decision Tree for anomaly detection using benchmark intrusion data, with performance assessed through training and test accuracy. On the whole, it can be concluded that each of the models demonstrated a high level of detection, with the best balance of strength and accuracy being demonstrated by Random Forest + SMOTE (train = 0.9997; test = 0.9969), which implies high generalization and a successful solution of the problem of a low percentage of classes. The Decision Tree was able to perform very well in test performance (train = 1.0000; test = 0.9929) but its score at perfect training indicates that there is a risk of overfitting (SVM = 0.9684; test = 0.9709); the performance was fairly lower but the results remain constant across the board. Taken together, these findings confirm the idea that ensemble-based models, specifically, Random Forest with imbalance management, are highly applicable to face one of the high-accuracy anomaly detection and enhance the ability to detect intrusions in IoT-like networks where the diversity of traffic and imbalance of attacks are frequent.

\subsection{Recommendations}
It is recommended to adopt Random Forest with SMOTE for anomaly detection where high accuracy and resilience are required, especially in imbalanced intrusion datasets. Moreover, imbalance in the classes must be addressed in a systematic way, using SMOTE or any other resampling techniques to equalize minority-class learning and minimize false negatives. If interpretability is of paramount importance, Decision Trees may be deployed, although pruning and validation controls should be utilized to minimize overfitting. In less compute-resource environments, or when maximum accuracy is not desired, SVM is also a viable option.  It is suggested to have good preprocessing, as it includes regular encoding, feature scaling where applicable, deletion of duplicates, and cautious handling of missing values and outliers to enhance the consistency of findings. Lastly, model explainability is to be implemented with SHAP or LIME to promote the trust and operational viability of cybersecurity decision-making.

\end{document}